# Air quality monitoring survey in german school classrooms during the COVID-19 pandemic 2021

Short title: Air quality monitoring in german classrooms

Meinhard Schilling[1], Simon Pelster[1], Timo Jelden[1], Eric Schlegel[2], Julius Mumme[2], Dean Ciric[2]
(1) TU Braunschweig, Institut für Elektrische Messtechnik und Grundlagen der Elektrotechnik, Hans-Sommer-Str. 66, 38106 Braunschweig, Germany
(2) fabmaker GmbH, Rebenring 33, 38106 Braunschweig, Germany

## Abstract (max 200 words)

The air quality in classrooms across all types of schools in Lower Saxony in the north of Germany has been monitored during June to December 2021 in a large study. Novel sensor instruments for air quality monitoring were wall mounted in 329 rooms in 50 schools and day nurseries. Each sensor instrument records data for carbon dioxide $CO_2$, sound level, temperature, and humidity. All collected data are transmitted for monitoring and further statistical evaluation by WiFi connection to the database. This study provides for the first time a detailed survey of the air quality situation and handling in schools during more than six months. The situation can be characterized by large variety in classroom area and volume combined with very diverse air ventilation possibilities. To control virus transmission via aerosol in classrooms an individual monitoring of air quality combined with high compliance for ventilation according to the signaled recommendation is urgently required.

6 Keywords: COVID-19, classroom monitoring, indoor air, window ventilation, CO2-concentration

Practical implications (5 sentences)
1. $CO_2$ monitoring for aerosol control is one of very few preventive measures against the COVID-19-pandemic, which allows quantitative, long-term, country-wide overview of effectiveness and compliance.
2. The air quality will be reliably kept in the hygienically acceptable range below 1000 ppm, if large, bright, for everybody well visible light signals are employed to signal the need for ventilation.
3. The compliance to such a visual signaling device for $CO_2$ monitoring is very high
4. The achieved air ventilation by window opening is sufficient for hygienically acceptable air in almost all situations in the monitored rooms independent from season and outdoor temperature, but static rules like 20 minutes alternating with 5 min ventilation are in most rooms not specific enough.
5. Critical rooms can be identified by evaluation of the stored data for further upgrading such rooms by forced ventilation or filtering

## 1. Introduction

Many proposals have been published during the current COVID-19 pandemic for suppressing further infections and ending the severe challenges for the health system [1, 2]. One main measure to reduce infections via aerosol transmission is a consequent exchange of indoor air, since the number of transmitted virus particles via aerosol exchange can be reduced significantly this way [3]. Compared to all other means this is one of the cheapest and ubiquitously available measures and therefore requires more public attention. For save teaching in all kind of schools and universities during the pandemic in many rooms air ventilation by opening the windows is the only available method for obtaining hygienically acceptable air since air filter units and air condition systems are generally not installed in classrooms in Germany. Beyond the actual pandemic situation since decades the uncontrolled and often very low air quality in classrooms has been investigated and discussed [4 - 7] and can lead to severely reduced learning success [8,9].

Infection models to describe and predict the possible effects of aerosol transmitted infections have already been published by many research groups [10, 11]. The emission of virus loaded aerosol by infected persons should be suppressed by all possible mechanical means, as there are masks, face shields and all kinds of mechanical barriers. In addition, the aerosol diffusion and spread should be suppressed by air management methods like ventilation, filtering and distance keeping. The human uptake of aerosol can be reduced by masks and the already mentioned mechanical barriers. The reduction of the number of contacts is an additional contribution, in its worst form as a lock-down phase.
For a long-term management of the pandemic the aerosol management is the only acceptable way for almost all affected persons to deal with the pandemic challenges, especially in our health system. Therefore, in this study we focus on the situation of air management by window ventilation to reduce the amount of virus loaded aerosols.

## 2. Materials and Methods

Starting in June 2021, new instruments "airooom" were continuously installed. For this study we analyzed the data of 50 schools and day nurseries in 329 rooms with a mean usage time of around 70 days of full 24 hours monitoring.
All rooms are located in the region around Hannover and Braunschweig in Lower Saxony.  The distribution of school types is shown in Figure 1. The 50 schools include 4 vocational training schools (Berufsschulen), 2 special schools (Förderschulen), 6 comprehensive schools (Gesamtschulen), 12 elementary schools (Grundschulen), 12 high schools (Gymnasien), 4 middle schools (Haupt und Realschulen), 6 day nurseries (Kindertagesstätten) and 4 other school types.

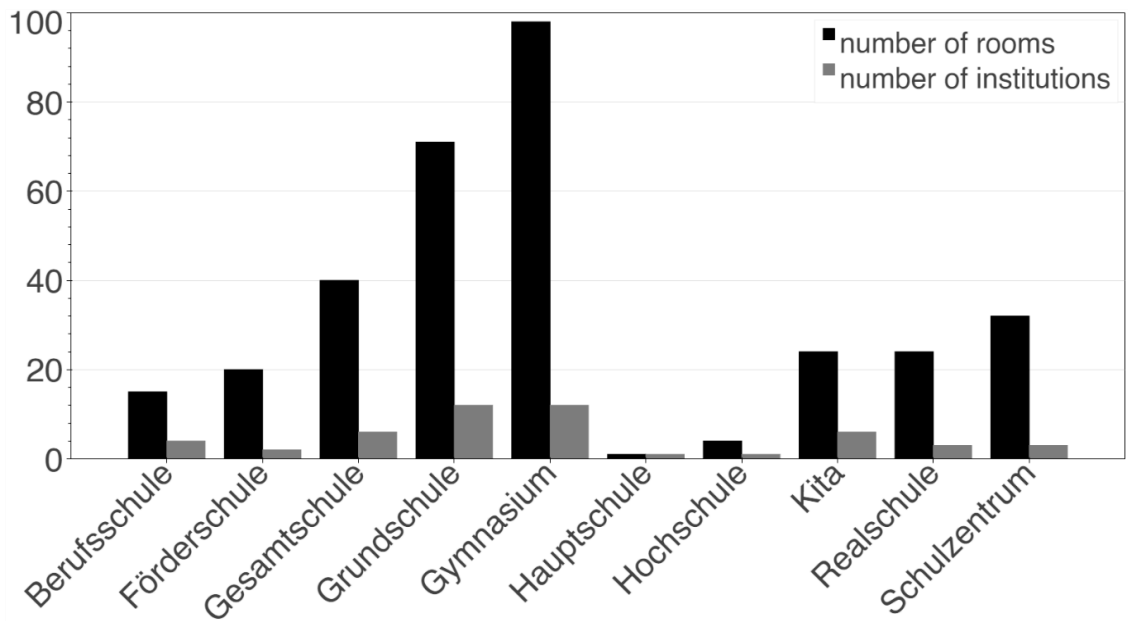

Figure 1 Number of schools and rooms for all school types.

This study focuses on school types for children and young adults up to the age of about 19. Further studies are needed to investigate the situation in universities and other institutions. Many results of this study are transferable nevertheless.

The geographical distribution of the monitored rooms is shown in summary in Figure 2.

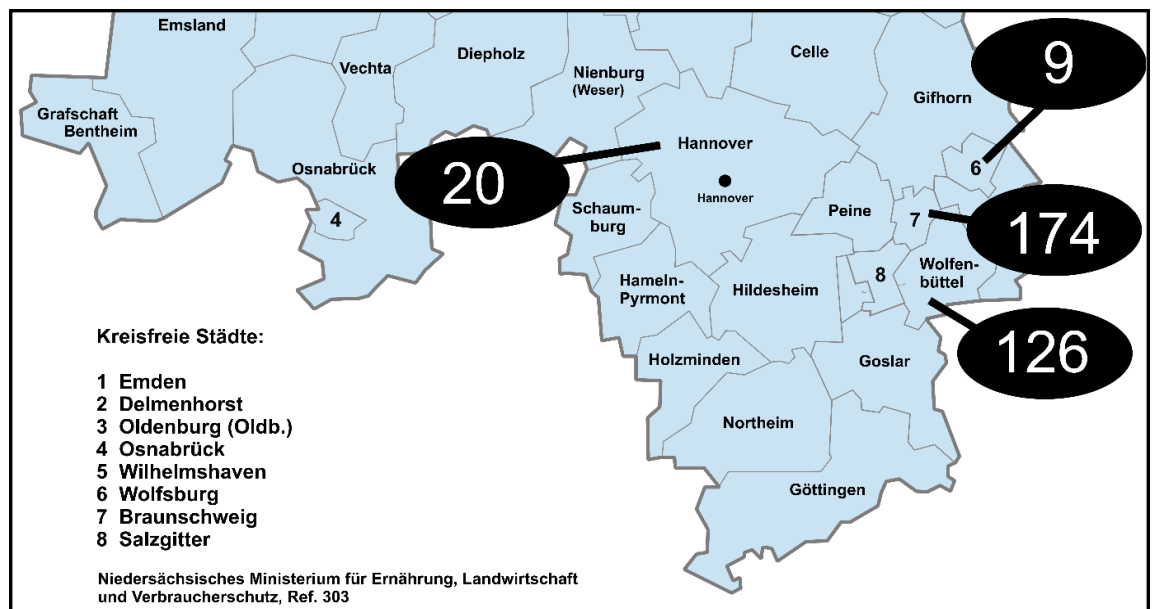

Figure 2 Geographical distribution of 329 monitored rooms in schools and day nurseries in the region around Hannover and Braunschweig in Lower Saxony [12].

The pandemic spreads wherever the virus can infect unprotected, not sufficiently immunized human beings. Thus, the infections in schools, where during the time of this study only very few children had been vaccinated, have contributed significantly to spreading the virus in society. Figure 3 visualizes the percentage of the COVID-19 infections in the age group 0 to 19 years to the total infections in Germany for each week as recorded since the start of the pandemic in the year 2019. The red line represents the percentage of the population occupied by this age group. It is visible that the younger portion of the population takes a significant part in the COVID-19 pandemic – especially since week 61 – which can't be explained by its fraction of the total population. About one fifth to one third of all infections occur in the age group up to 19 years since week 61 in march 2021. Since this time the percentage of 18,4 % of this age group of the total population (red line in Figure 3) is continuously exceeded by the percentage of incidences in this age group compared to the total incidence of the whole population.

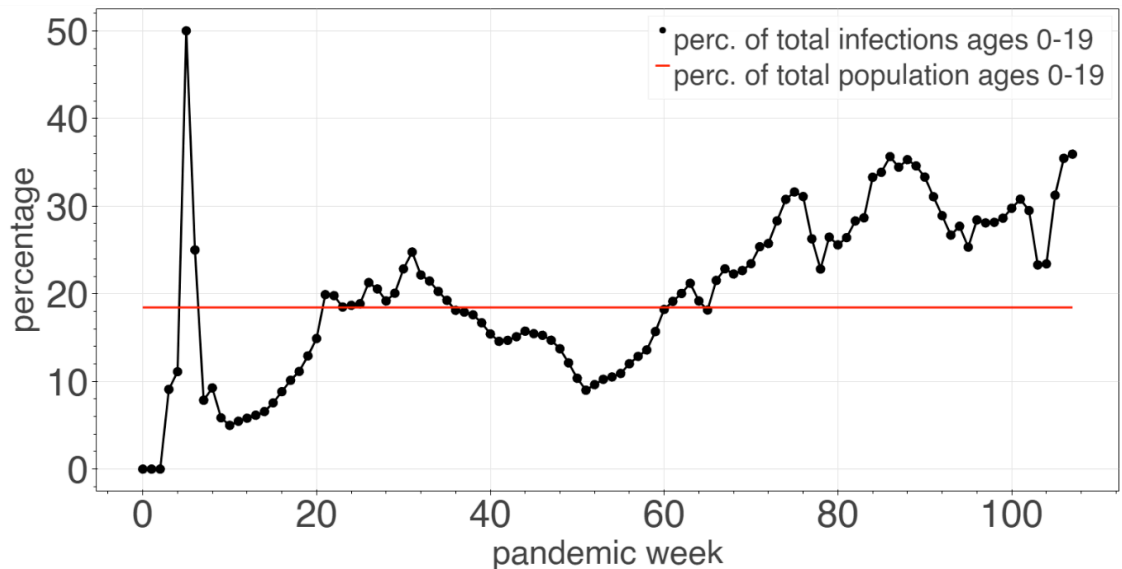

Figure 3 Percentage of infections in age group 0-19 years to total infections since the start of the pandemic in Germany [13, 14].

**Room Situation**

In our study we investigated the effectiveness of measures against the pandemic in the monitored school classrooms and day nurseries. For this purpose, we recorded the spread of classroom sizes and ventilation possibilities for the monitored classrooms. The sizes for a nominal class size of up to 30 pupils vary between 35 m² and 140 m² with corresponding room volumes of 105 m³ and 416 m³ depending on the different room heights. The ventilation in these rooms is only possible by opening the windows. These windows also have a wide variety of area between 0.25 m² and 4 m² and are not all able to be fully opened. This variety of classroom situations in this study is depicted in Figure 4.

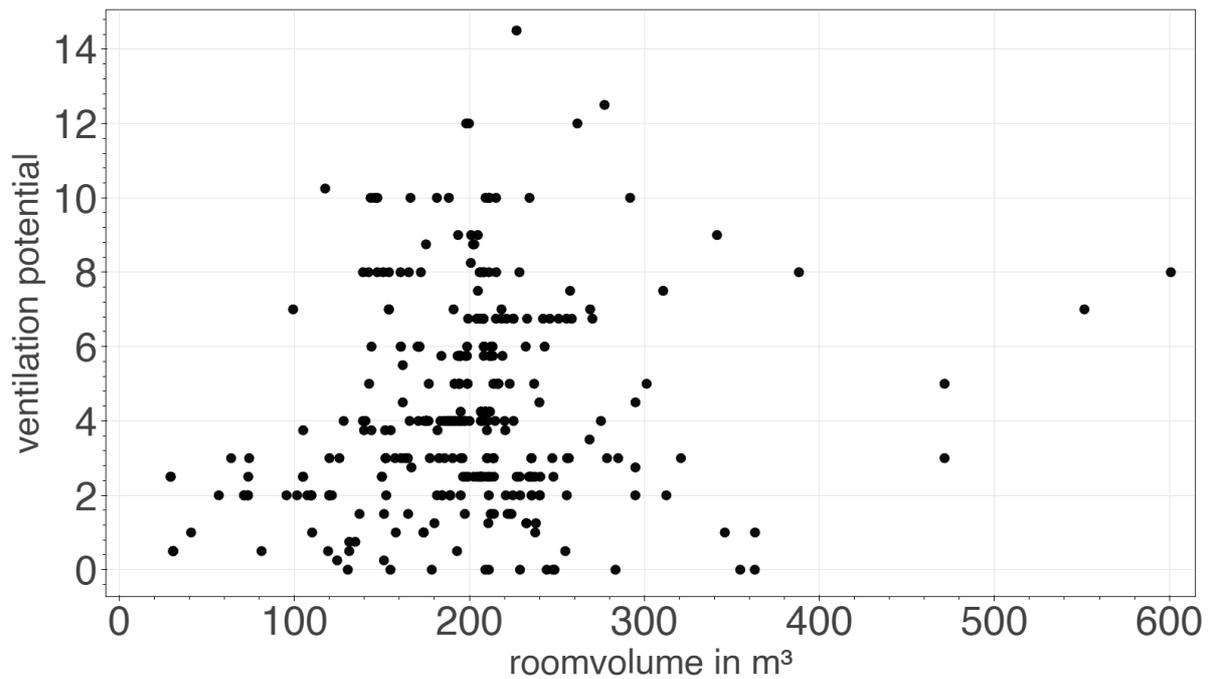

Figure 4 Ventilation situation in all 329 monitored classrooms in Lower Saxony.

The ventilation potential we define as the number of windows, which can be fully opened plus the number of partially openable windows weighted with a factor of 0.25. The distribution of the median, mean and 95% percentile of the room volumes on the different school types is depicted in Figure 5. We monitor the largest room volumes per pupil in high schools (Gymnasium) and the lowest in special schools (Förderschule) differing by almost a factor of two. The Umweltbundesamt in Germany recommends more than 7.5 m³ per person in schoolrooms [15]. This is approximately the situation as depicted in Figure 6.

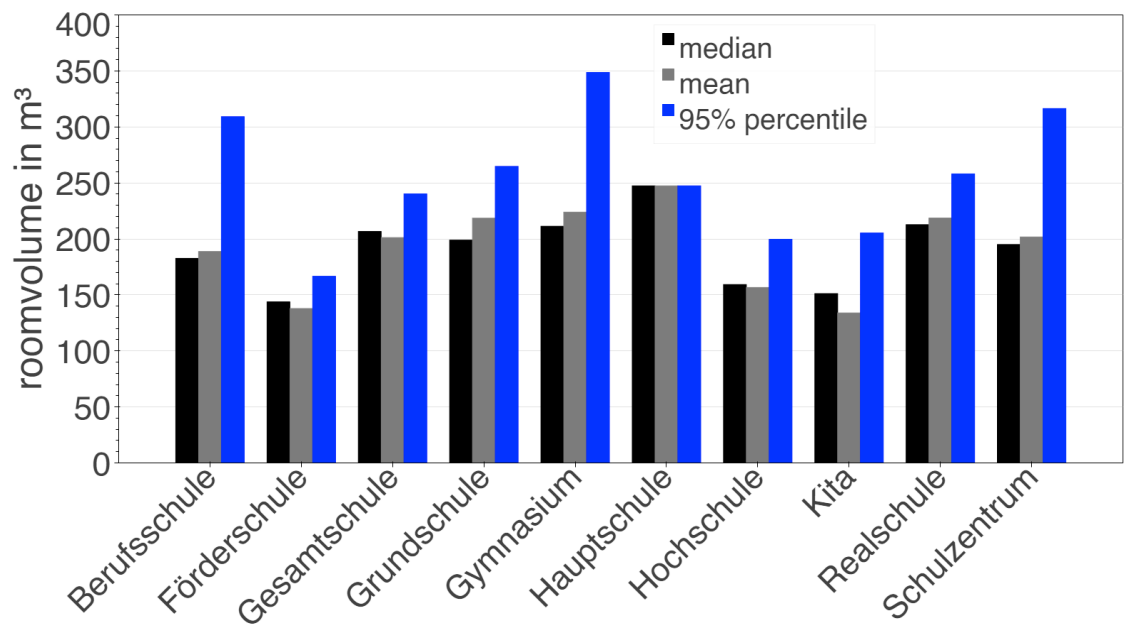

Figure 5 Median, mean and 95% percentile of room volume for each school form

Figure 6 shows the room volume divided by the maximum number of occupants who use the room for each school type. The maximum occupancy for each room was determined by counting the chairs in each classroom when installing the sensor instrument.

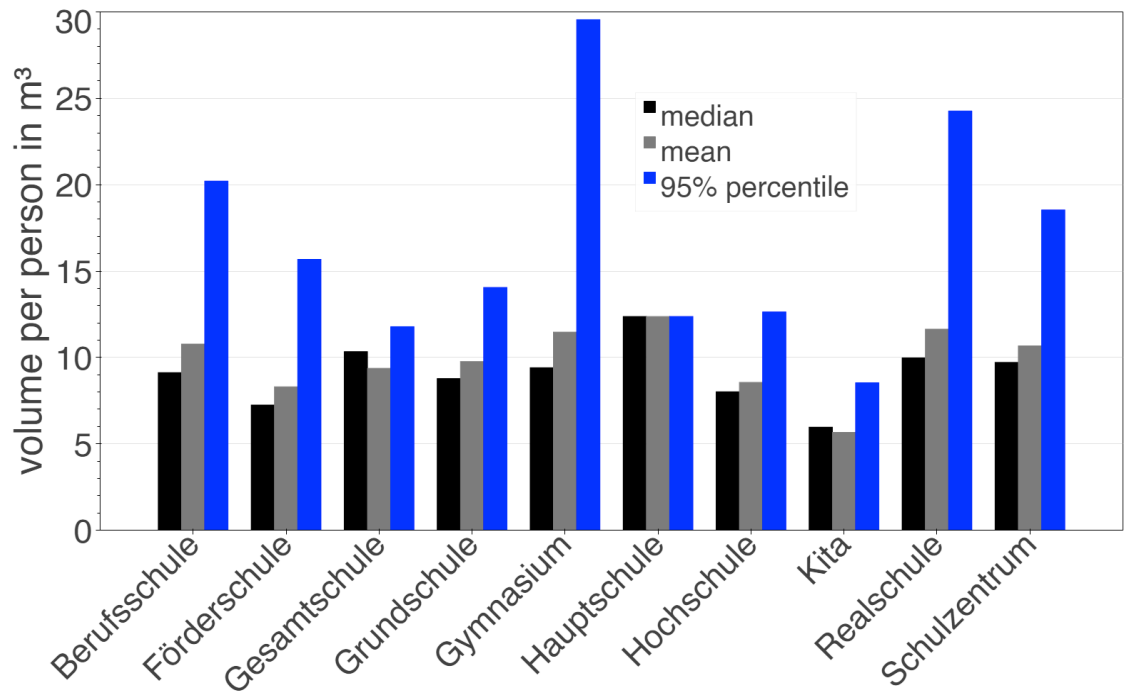

Figure 6 Median, mean and 95% percentile of room volume per person for each school form

The official measures of the government of Lower Saxony against the pandemic during 2021 supported the installation of mobile filter units in the classroom for the rooms with worse ventilation situation than the monitored rooms in this study. As an alternative the possible installation of window ventilation systems is discussed, which at the time of this publication has not been realized to a larger extend, and the use of $CO_2$-sensor instruments for monitoring the air quality during the school lessons. Most of the available mobile $CO_2$ sensors provide no data transmission, have no data storage system attached and thus no data evaluation is possible. This was the main motivation to develop the new instrument "airooom" of this study, which is permanently installed and has data transmission capability.

### Sensor instrument „airooom"

This was the background for our study with the sensor instruments "airooom", which were officially supported and installed in 329 rooms in Lower Saxony and made it possible to collect the data evaluated in this study.
This sensor instrument consists of a traffic light like case of 30 cm height containing three colored large, bright signal LED lights in green, yellow and red as depicted in Figure 7. The colors signal the necessity for air exchange: green signals a $CO_2$ concentration lower than 800 ppm, red a $CO_2$ concentration above 1000 ppm and yellow the transition between these states between 800 ppm and 1000 ppm. The value of 1000 ppm is regarded as hygienically acceptable [16] and recommended in Lower Saxony for school classrooms.

The possibility of direct monitoring of aerosol concentration has been investigated, but turned out to be too expensive and too error prone to be included in the measurements. The simple aerosol detection sensors are not suited to distinguish air dust particles from aerosol droplets and thus the measured aerosol content of the air seems to increase significantly if the windows are opened due to fine dust particles from the environment or traffic entering the room. This effect makes it very hard to distinguish fine dust from the emitted aerosol of the human beings in the classroom.

In addition to the $CO_2$ sensor, the instrument contains a microphone for averaged sound level measurement without any possibility to record spoken words, a thermometer, a humidity sensor and the microcontroller system for data collection and transmission of the data via WiFi. The transmitted data are collected in a database for evaluation.

The monitoring of all sensors by the transmitted data allows it to verify the correct sensor operation of all instruments. The return during the night to the atmospheric value of 412.5 ppm $CO_2$ concentration [17] at the time of publication provides a good control of possible long-term drift effects in the $CO_2$ sensors. The same sensor monitoring is accomplished for the temperature, humidity and sound level sensors as well as for the WiFi connection and power supply of the instruments. Fault conditions are detected with below minute delay and can be repaired on the same day. The position for mounting the instrument in the classroom was selected more than 1 meter from room corners, doors or windows, not above furniture, not above human beings and about 1.8 m above the floor level. The access to electricity should be possible with short cords to prevent accidents.

This sensor instrument thus provides 3 levels of evaluation: Direct feedback to the teacher for learning situation adopted air ventilation, monthly feedback for school operator administration about the room situation, and data collection for further statistical evaluation as in this first study.

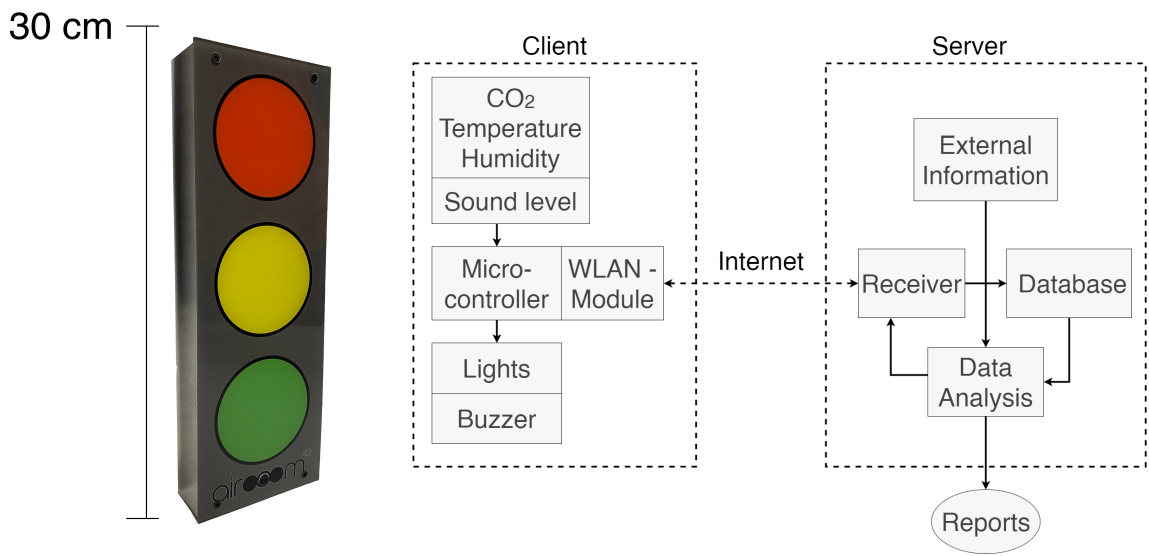

Figure 7 The sensor instrument "airooom" consisting of 3 large, bright signal LED lights, sensors for air quality and wireless data transmission by its microcontroller.

## Methods

In order to analyze the ventilation situation from the monitored data we need to obtain the air exchange rates during different ventilation scenarios. Extracting the air exchange rate is only possible when the number of occupants is known. But this number can change drastically over time especially during the pandemic since some schools divided their classes in two groups in times with high local infections. This fact generates the need for extracting the number of occupants from our data dynamically.
To comply with these requirements and to estimate the number of occupants and the air exchange rate at a given time we assume the following model based on [18, 19] for the $CO_2$ concentration in the rooms.

$$V \cdot \frac{\partial C_i}{\partial t} = -n \cdot V \cdot C_i + n \cdot V \cdot C_a + S \qquad (1)$$

$C_i$: $CO_2$ concentration in the room
$C_a$: $CO_2$ concentration outdoors
V: room volume, [V] = L
n: air exchange rate, [n] = 1/h
S: $CO_2$ generation rate, [S] = L/h
t: time, [t]=s

In Eq. 1 we assume the air exchange rate and $CO_2$ generation rate to be constant over time.
The concentration $C_a$ denotes the $CO_2$ background concentration of the atmosphere ($C_a$ = 412,5 ppm at the time of publication). When calculating the time dependent $CO_2$ concentration an integration constant has to be added, which is identified as $C_{i,0}$, the concentration offset present at the start of the measurement. Changes in number of persons and the opening of windows have to be detected to find suitable time frames for analyzing.

### Occupancy detection

Since the occupancy of most of the rooms follows a classical alternation of teaching time and breaks, we assume the number of persons in the room to fluctuate between the full class size and zero with no transitioning states. A combination of the sound level and trend of the $CO_2$ concentration is used to classify the occupancy times: Rising $CO_2$ concentration is assumed to correlate with human occupancy of the room. In order to distinguish from fluctuations in case of doubt (small increase rate, or short term rising) the sound level is evaluated in addition.

Figure 8 shows an example result of our algorithm in a room over a school day. The time spans during which the occupancy is classified as "class present" are colored in green based on excess noise level above the background noise level together with an increasing $CO_2$ concentration due to the presence of human beings. The sound level alone would be a too weak criterion for occupancy since street noise could have the same effect. The background noise level is detected during decreasing or constant $CO_2$ level. The minimum duration for each state is set to 5 minutes to avoid misclassification of quiet times during the lessons.

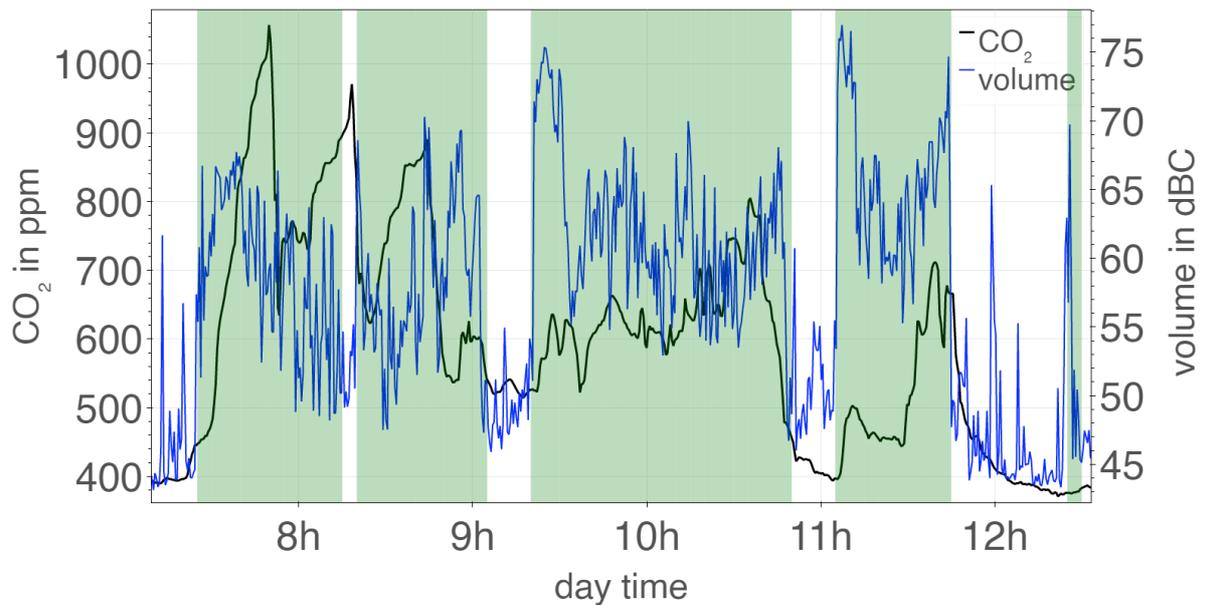

Figure 8 Occupancy detection example with detected occupancy marked in green areas.

Time intervals of decreasing $CO_2$ concentration on the other hand can be used to identify air ventilation periods. An example is shown in Figure 9. The blue areas mark the recognized ventilation periods.

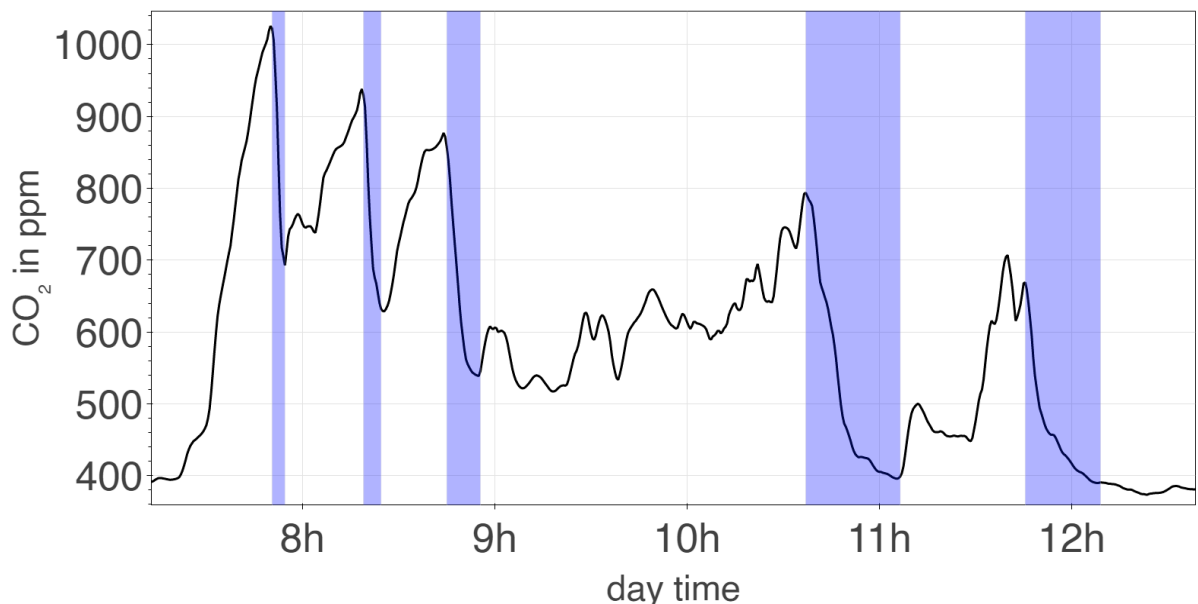

Figure 9 Example of ventilation recognition in the same classroom as investigated in Fig. 8. Low pass filtered data are used to simplify the ventilation recognition.

We only identify significant drops in $CO_2$ concentration as ventilation periods since we assume that in these situations active measures were taken by the occupants to ventilate the room (e.g. opening windows). Furthermore, we assume that these ventilation periods produce a constant air exchange rate with a constant number of windows opened.

### Number of persons

To estimate the number of persons during occupancy times we assume an air exchange rate of zero, which leads to a linear rise in $CO_2$ concentration. The minimum air exchange rate occurs when all the ventilation possibilities (e.g. windows and doors) are closed. To identify such a situation, we search for the steepest stable rise in $CO_2$ concentration over a full day and assume that a linearization of the increasing $CO_2$ concentration dependence on time is justified.

Next, we estimate the mean $CO_2$ generation rate per occupant using Eq. 2 according to [20].

$$\dot{V}_{CO_2} = BMR \cdot M \cdot T \cdot 0.000179 / P \tag{2}$$

$\dot{V}_{CO_2}$: $CO_2$ generation of an occupant, $[\dot{V}_{CO_2}]$ = L/s
BMR: basal metabolic rate, [BMR] = MJ/day
M: level of physical activity [20]
T: air temperature, [T] = K
P: pressure (102 kPa), [P] = kPa

The BMR can be estimated based on the age and weight of the occupants [21, 22]. M can be estimated based on the expected activity and is set to a value corresponding to "sitting reading, writing, typing" [23].

Now we calculate the number of occupants N in Eq. 3 by analyzing the $CO_2$ rise ($C_{t2} - C_{t1}$) in a time span between $t_2$ and $t_1$ with:

$$N = V(C_{t_2} - C_{t_1}) / ((t_2 - t_1) \cdot \dot{V}_{CO_2}) \tag{3}$$

N: Number of occupants
V: room volume, [V] = L
$C_{ti}$: $CO_2$ concentration in the room at times $t_1$, $t_2$
$V_{CO2}$: $CO_2$ generation of an occupant, $[V_{CO2}]$ = L/s
t: time, [t]=s

### Air exchange rate

The decrease of the monitored $CO_2$ rate allows to estimate the air exchange rate during a ventilation process identified by the method described before. This might change each time, since the number of opened windows and the way they are opened might be different.

To estimate the air exchange rate n, we need to solve Eq. 1 with n > 0 [18], which leads to Eq. 4.

$$C_i(t) = (C_{i,0} - C_a - \frac{S}{n \cdot V}) \exp(-nt) + \left(C_a + \frac{S}{n \cdot V}\right) \tag{4}$$

V: room volume, [V] = L
n: air exchange rate, [n] = 1/h

S: $CO_2$ generation rate, [S] = L/h
$C_{i,0}$: $CO_2$ concentration offset in the room at the start of the measurement
$C_a$: $CO_2$ concentration outdoors

With that knowledge we can derive the air exchange rate n for time segments where S is constant during ventilation periods by fitting the model from Eq. 4 to real world data using non-linear least squares.

For the ventilation periods in Figure 9 the following air exchange rates were evaluated (counted from left). Assuming the occupancy to be constant over the ventilation period does not always hold perfectly as you can see in ventilation 4. This fourth ventilation falls into a time interval at the beginning of a break when pupils are about to leave the room as can be detected from the sound data in Figure 8. Since more than 50 % of the ventilation time the pupils already left the room we classify this ventilation as one of an unoccupied room in Table 1. But since this assumption simplifies the model and the example ventilation 4 presents the worst-case scenario we still applied this constraint.

| Ventilation | 1 | 2 | 3 | 4 | 5 |
|---|---|---|---|---|---|
| Air exchange rate in 1/h | 18 | 17 | 16 | 6 | 11 |
| Persons present | 34 | 34 | 34 | 0 | 0 |

Table 1 Example air exchange rates derived from the model

## 3. Results and Discussions

In this section we present the evaluated results in two subsections: *Statistics of air ventilation* explains the environmental situation and in *Compliance overview* we explain how much the persons responsible for air ventilation followed the recommendations of the sensor instrument signals.

### Statistics of air ventilation

Using the ventilation detection algorithm, we identified and evaluated in total over 50,000 ventilation events of which 30.6 % took place without the school class being present (during breaks or after class) and 50.6 % with the class present. In 18.8 % of ventilations the occupancy couldn't be decided and these events were excluded from further evaluation. By this procedure random, ambiguous events were removed from the statistics without creating a bias for results where occupancy played a role. We found that very often rooms were ventilated at the end of the school day. The median ventilation duration was 8 minutes and the median number of ventilations per day was

4. Only such days were evaluated when the rooms were occupied long enough that at least one ventilation event was identified.

The mean $CO_2$-concentration during times when the rooms were occupied was 620 ppm with a median of 595 ppm. The 95% percentile $CO_2$-concentration was 945 ppm. We found that only in 0.94% of school days the mean $CO_2$ concentration exceeded 1000 ppm. In 30.74% of the monitored school days – mainly during the winter months – the $CO_2$ concentration exceeded the 1000 ppm mark at least once leading to a red signal light.

The ventilations in all of the monitored rooms were collected in Figure 10. The y-axis marks the total duration of each ventilation, while the x-axis marks the time difference to the last ventilation in the same room. The circle marks the point, which corresponds to the general ventilation recommendation of "20-5-20", which advises a ventilation duration of 5 minutes in 20 minutes intervals [24]. It is obvious that this static recommendation is not suited to control the air situation in a wide variety of rooms. We find that the true ventilation time is often much longer or the intervals between succeeding ventilations are much shorter in our monitored rooms. A recommendation of "5-15-5-15-5" with 5 minutes ventilation followed by 15 minutes of closed windows seems to reflect the data based recommendation of the sensor signals much more appropriately for the investigated rooms in this study. For these rooms with restricted ventilation possibilities the 20-5-20 rule seems not suitable and might lead to too high $CO_2$ values during the course of the school hour.

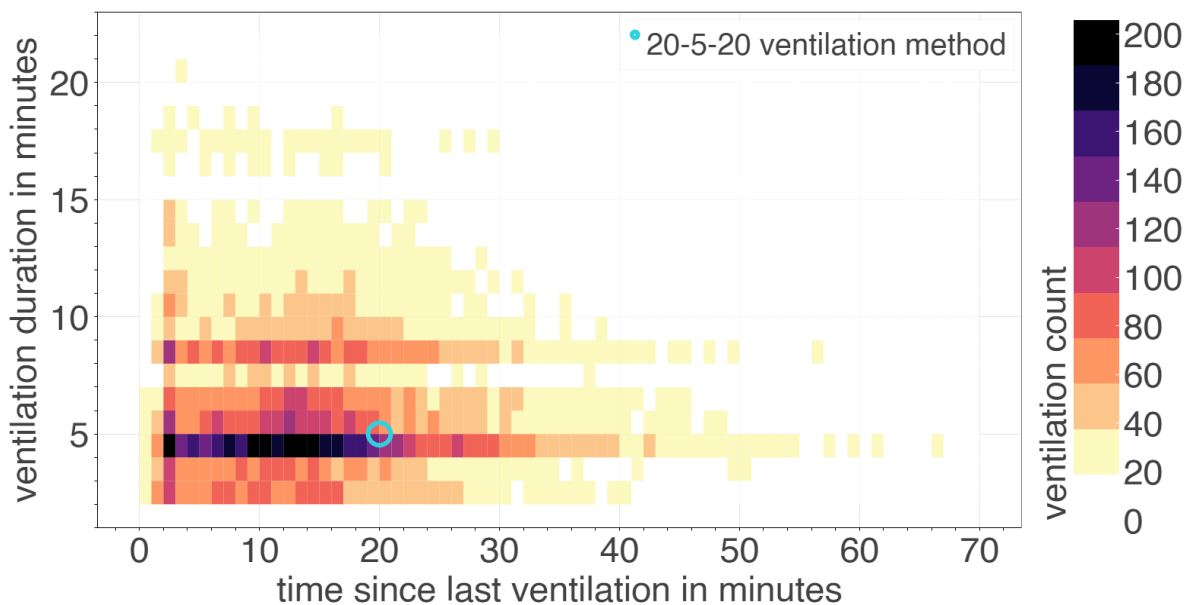

Figure 10 Ventilation duration and time difference to the last ventilation in the monitored rooms. The circle marks the general recommendation of "20-5-20".

The state changes of the sensor instruments traffic light signal can always be determined by analyzing the values of $CO_2$ concentration. As a next step, we analyzed the statistics of ventilations in conjunction with state transitions of the light signal. Figure 11 shows the distribution of state changes for all ventilation periods that took place with the classes being present. First of all, we found many air ventilation events where the light signal state didn't change during the ventilation period. In the case of a ventilation starting in the "green" light state and ending in the "green" light state, obviously the windows were opened voluntarily without signal change from the instrument. These ventilations are termed "green-green" in Figure 11. Additionally, there are as well "red-red" and "yellow-yellow" ventilation events, where the ventilations started during a "yellow" or "red" warning state, but the windows were not kept open until the $CO_2$ level dropped below 800 ppm and the lights didn't return to the "green" state. These events ("red-red", yellow-yellow" and "red-yellow") were rarely observed and the ventilation might have been stopped for some organizational reasons. The majority of ventilation events started with the transition into "yellow" or "red" state and the windows were opened until the lights returned into the "green" state and the $CO_2$ value dropped below 800 ppm.

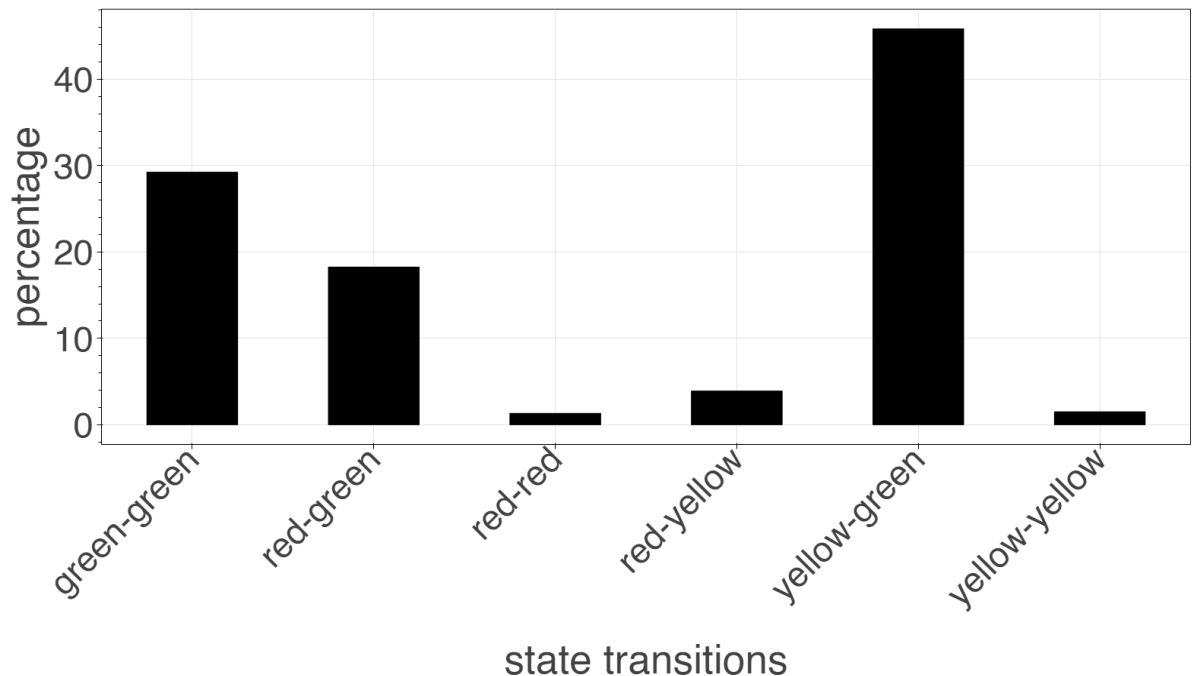

Figure 11 Percentage of ventilation events, when the monitored rooms were occupied, categorized by the start and end state of the sensor instruments traffic light signals (25571 ventilation periods combined).

The instruction given to the teachers after installing the "airooom" instruments had been to open windows when the signal light turns red, but we found that a high amount of ventilation already began while the traffic light was yellow. Furthermore, we see that 94,42 % of ventilations starting with red ended with yellow or green, which indicates that in almost all cases the occupants were successful in changing the room air quality back to an acceptable level.

In Figure 12 it is shown that the median and mean of the $CO_2$ concentration during the occupied time in the classroom stayed below 800 ppm for all evaluated ventilation events during the whole monitored time in nearly all of the school forms. The 1000 ppm mark was only exceeded by the 95 % percentile of the school form "Realschule". This is a very important result of this study, which convincingly demonstrates how effective the novel sensor instrument with large area LED-light signals can support the effective ventilation for hygienically acceptable air quality by window opening in all monitored rooms independent of size, volume and ventilation potential (number of windows, area of windows).

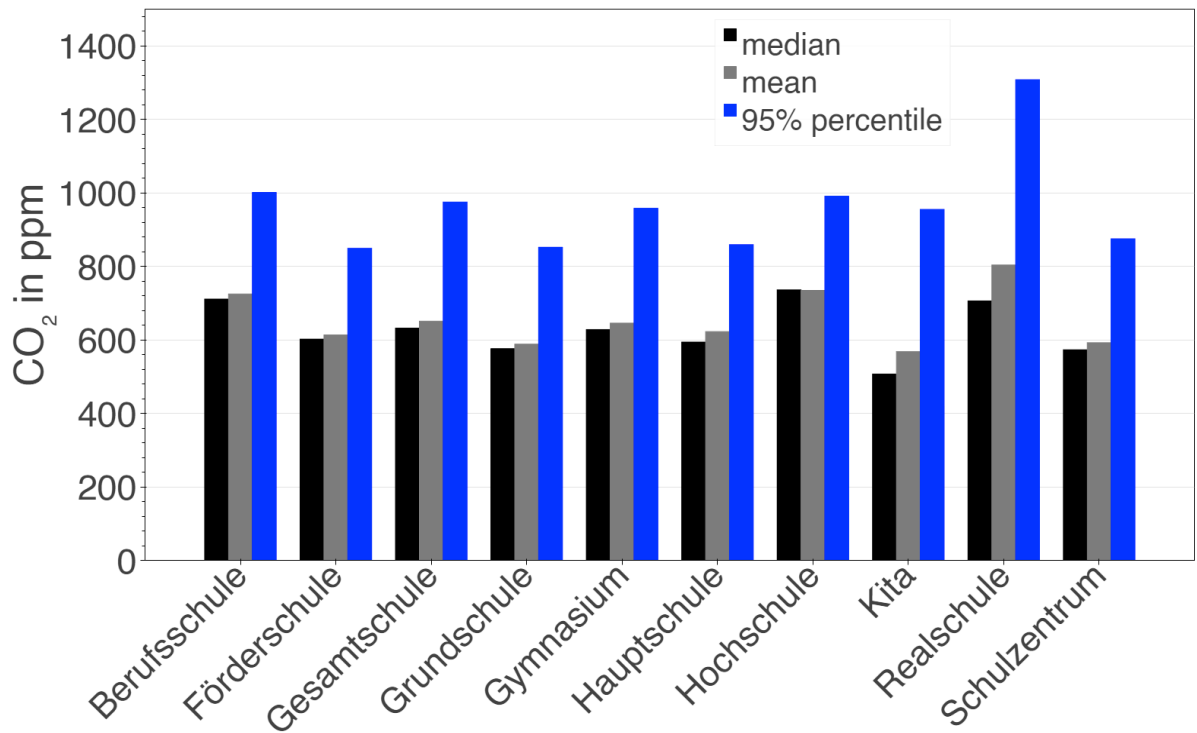

Figure 12 Median, mean and 95% percentile $CO_2$ concentration for each school form

We can also evaluate the $CO_2$ distribution for each room by calculating the 5%, 50% and 95 % percentiles for the occupied times, which is shown in Figure 13. Only rooms with over 5000 available data points were considered. Ten sensor instruments in rooms with 5 % percentile below 300 ppm were dismissed from further evaluation due to less accurate calibration or sensor offsets. These instruments were replaced or recalibrated on short notice. No room median exceeded the 1000 ppm mark and even the 95% percentiles stayed below 1000 ppm in 77 % of rooms.

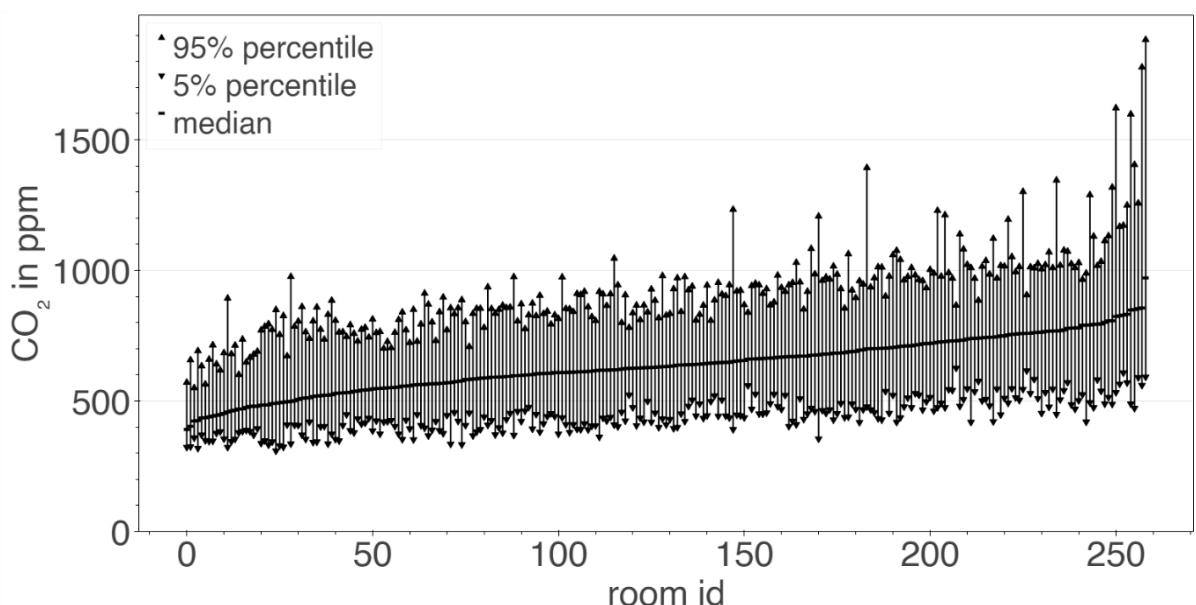

Figure 13 Median, 5% and 95% percentile of $CO_2$ concentration in analyzed rooms

**Influence of temperature on air ventilation**

In all schools concerns arose, how air exchange by window ventilation can be managed during the winter months with low outdoor temperatures. The adaption of ventilation time and ventilation intervals to the room conditions reduces the temperature reduction to the minimum required measures to keep hygienically acceptable air quality. Therefore, we investigated how the outdoor temperature has influenced the air ventilation events.
In Figure 14 the sensor instruments state changes for ventilation with an outdoor air temperature of T < 3 °C is visualized alongside to the state changes with T > 14 °C. The percentage of ventilations for each state transition is shown for both temperature ranges.
The outdoor temperatures used in this paper were downloaded from the official site of the german weather service ("Deutscher Wetterdienst") [25]. Weather stations located in the same county as the corresponding school were used.

Comparing the percentages in each temperature group for a given state transition reveals interesting results. The transition "green – green" of voluntary ventilation events is very common with warmer outdoor temperatures (56 %) while it is occurring less often when it's colder outside (20 %). The "red – green" transition happens significantly more often (24 %) in the colder temperature group compared to the warmer group (6 %).
The "red – yellow", "red – red" and "yellow – yellow" transitions are rare in both temperature groups but occur more frequently with colder outdoor temperatures.
Lastly the "yellow – green" transition is common for both temperature ranges, but the ventilations belonging to colder temperatures are again more frequent.
This result suggests that with lower outdoor air temperatures the desire to ventilate without a yellow or red light signal of the sensor instrument decreases. Subsequently, the amount of time the "airooom" signals red or yellow increases. This reflects the problem with decreasing average temperature during school hours in the classroom, if the air ventilation situation requires too frequent ventilation. Nevertheless, in our monitored rooms the compliance to the warning signal of the "airooom" instrument was still excellent, even when outdoor temperature was low during the winter months.

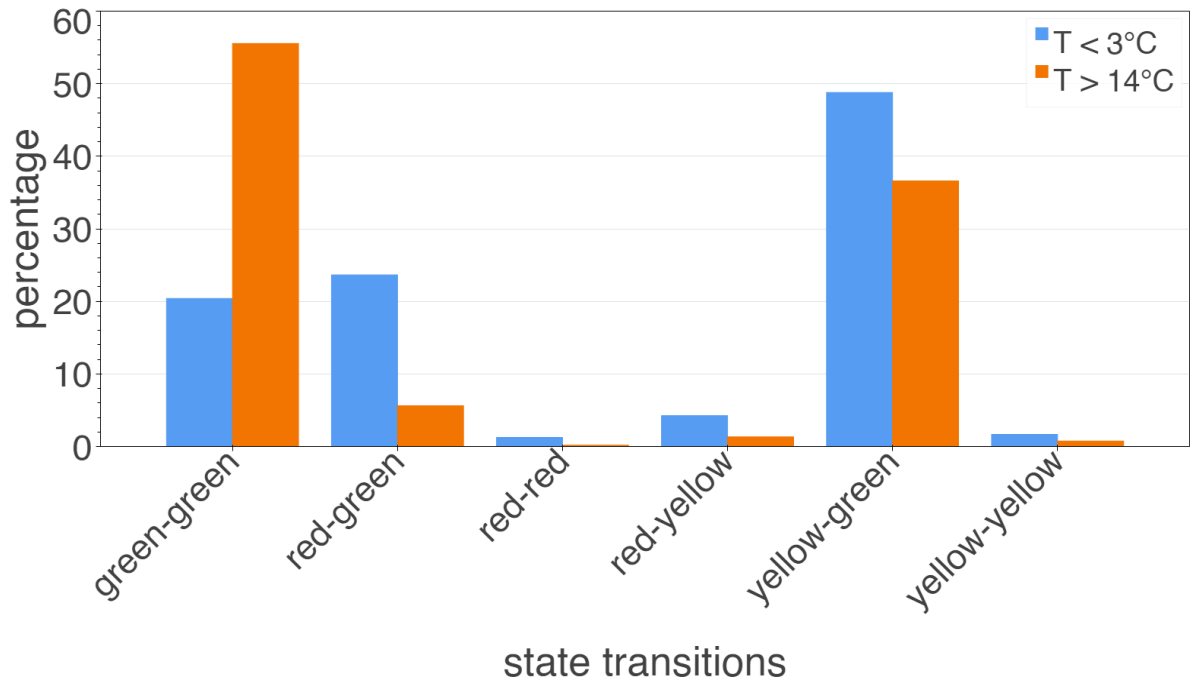

Figure14 Ventilation events for low (5294 ventilations combined) and high temperatures (3475 ventilations combined)

To analyze the influence of the outdoor temperature on the number of ventilations while the sensor instruments signaled "green", we visualized the percentage of these ventilations over the outdoor temperature in Figure 15. The number of ventilations while the signal was "green" is shown with a green line, while the rest of the ventilations is shown with a red line.

It can be seen that the percentage of ventilations with a green sensor instrument signal rise with the outdoor temperature. At 15 °C and above more than 50% of ventilations are performed voluntarily without the sensor instrument signaling "yellow" or "red".

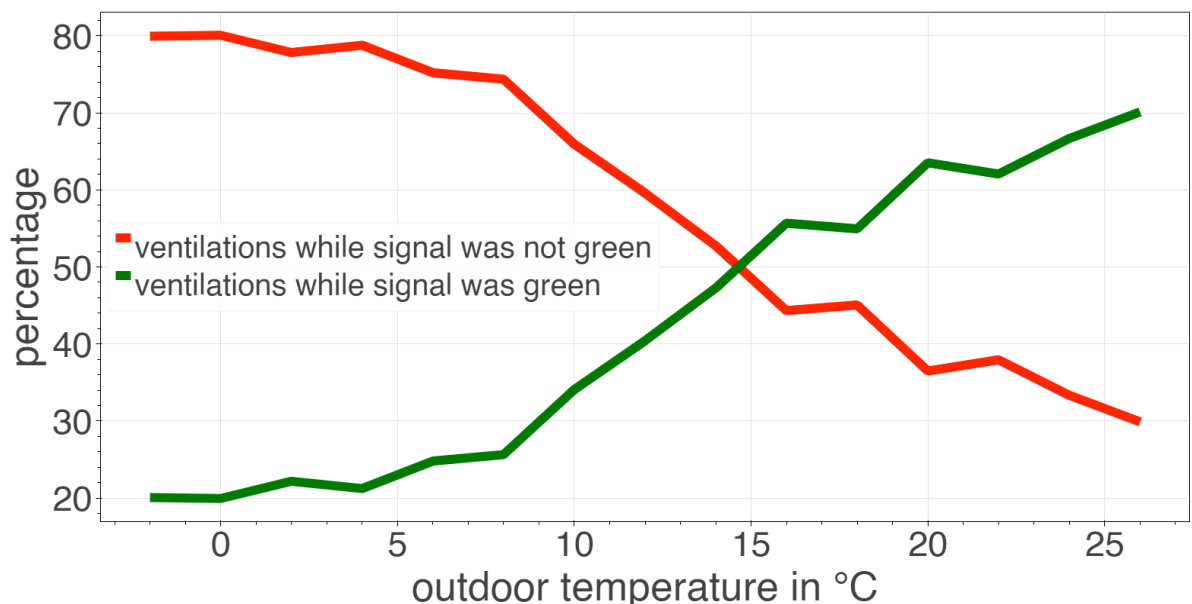

Figure 15 Percentage of ventilations belonging to "green – green" voluntary ventilation events compared to the rest of ventilation events

Despite the increasing number of red phases at lower temperatures the duration of the red phase remained of same duration independent from the outdoor temperature, as can be seen in Figure 16. Each temperature bin is 2 °C wide and is based on at least 100 red phases. The total number of analyzed red phases amounts to 12212. The increase of red-phase duration for high outdoor temperature we attribute to the lower air exchange rates when the indoor and outdoor temperatures are about the same. Longer compliance times in summer may contribute here as well.

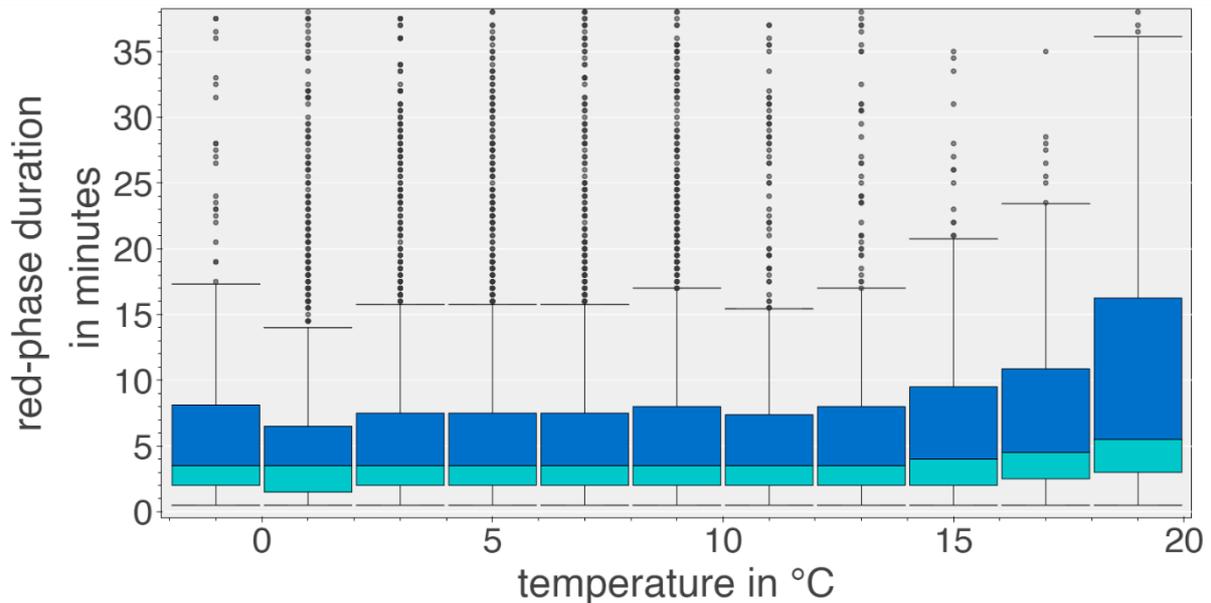

Figure 16 Boxplot of the duration of the red-phase in dependence of the outdoor temperature in °C

The temperature influence on the ventilation behavior can be evaluated additionally by comparing the heatmaps in Figure 17 and Figure 18. The ventilations in these Figures belong to the outdoor temperatures T > 14 °C and T < 3 °C, respectively. For both Figures ventilation durations up to 25 minutes and times differences to the last ventilation up to 70 minutes were considered. The time since last ventilation spans wider for warmer temperatures. Also, the ventilation situation in colder temperatures is much denser below 5 minutes ventilation duration.

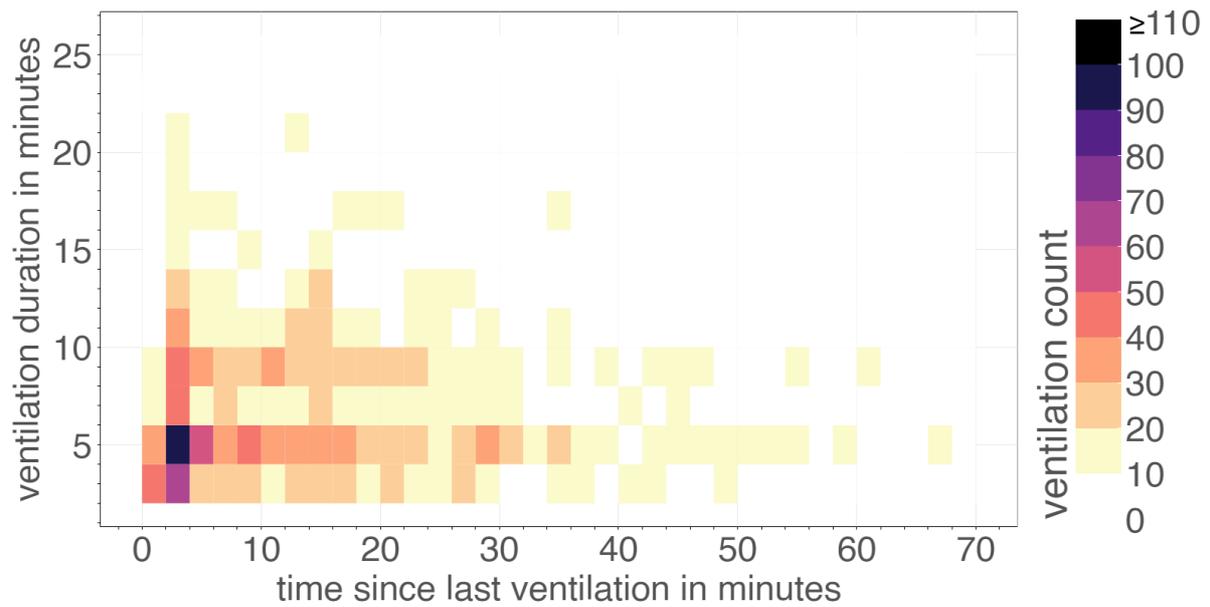

Figure 17 Ventilation duration vs. time since last ventilation for the outdoor temperature distribution with T > 14°C (3626 ventilation periods visualized)

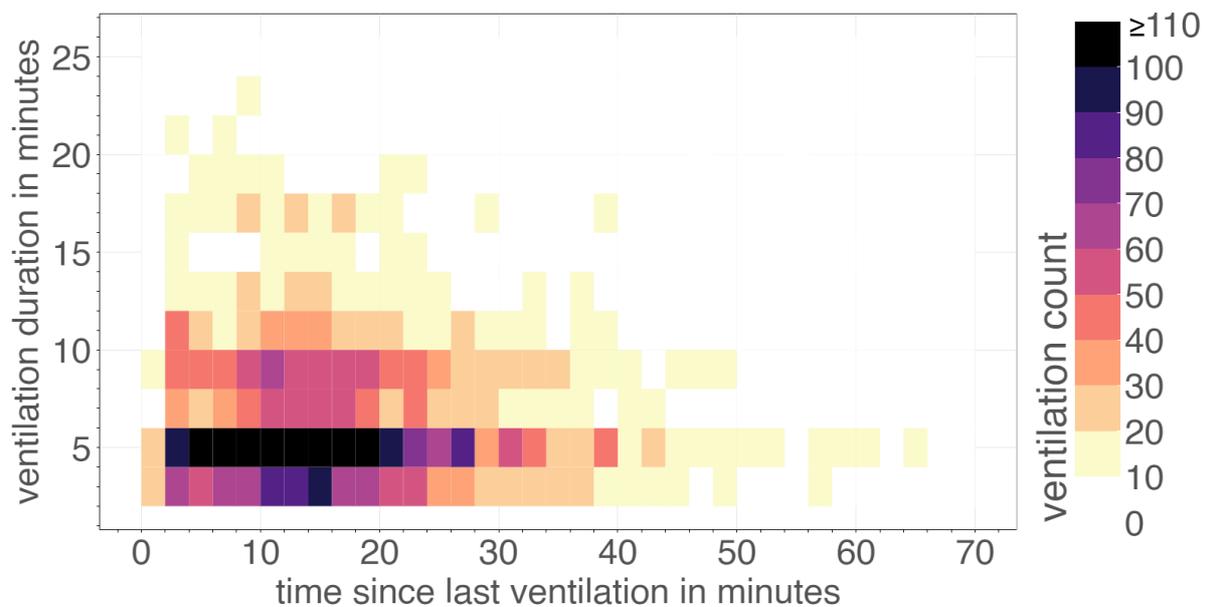

Figure 18 Ventilation duration vs. time since last ventilation for the outdoor temperature distribution with T < 3°C (6499 ventilation periods visualized)

It would be of great interest to the users in the monitored rooms (pupils and teachers) to improve the air quality, while keeping the air temperature at an acceptable level. Figure 19 compares the outdoor air temperature with the indoor air temperature in the analyzed rooms. The outdoor temperatures were grouped in 2 °C wide bins and each bin contains at least 10,000 data points. The 25 % percentiles of the indoor temperatures barely ever fall below 18 °C, which can be classified still regarding thermal comfort as "mostly comfortable" temperature level assuming the relative humidity to be in range between 32 % and 86 % [26].

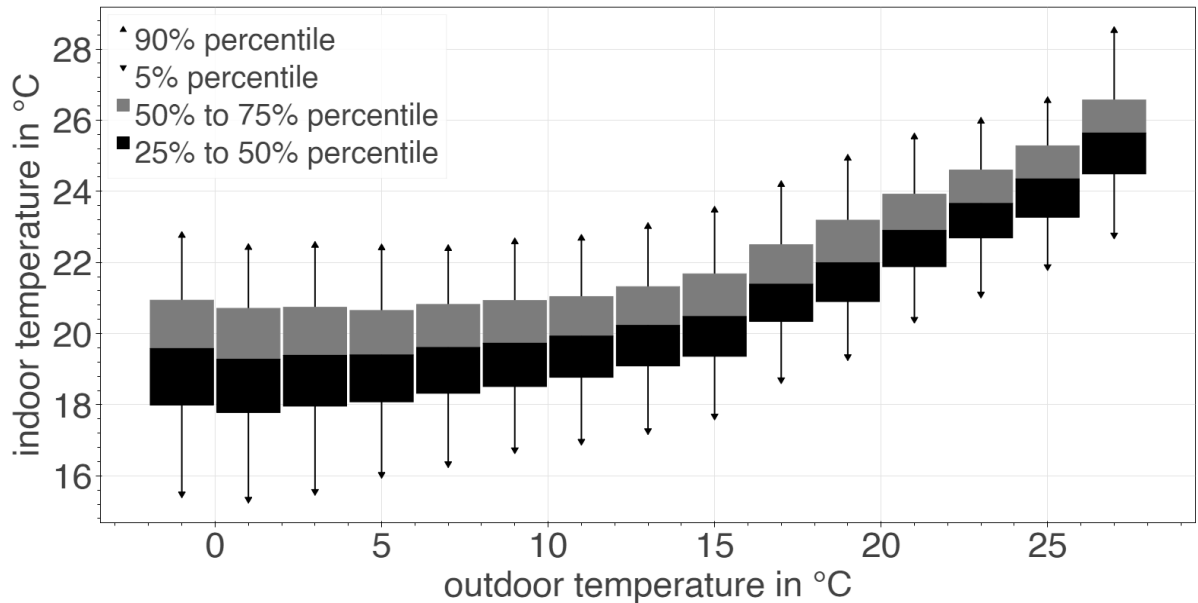

Figure 19 Comparison of outdoor and indoor air temperature.

**Influence of number of windows on ventilation**

Multiplying the air exchange rate in 1/h with the room volume results in the air flow which is independent from the room volume. In Figure 20 the air exchange rate in 1/h and the corresponding air flow in m³/h is plotted against the ventilation potential, defined as the number of windows, which effectively contribute to ventilation. Only rooms without technical ventilation systems installed were selected and every data point is based on at least 200 ventilations. Furthermore, only ventilations of occupied rooms were considered. We find no correlation between the air exchange rate and ventilation potential ($R^2$=0.0 for linear regression of median values). This could indicate that the occupants not necessarily open all available windows when ventilating. But it is also possible that the air exchange rate is limited by a different factor, e. g. to limit ventilation to a maximum air speed at a door of the classroom according to protest of the pupils sitting closer to the doors.

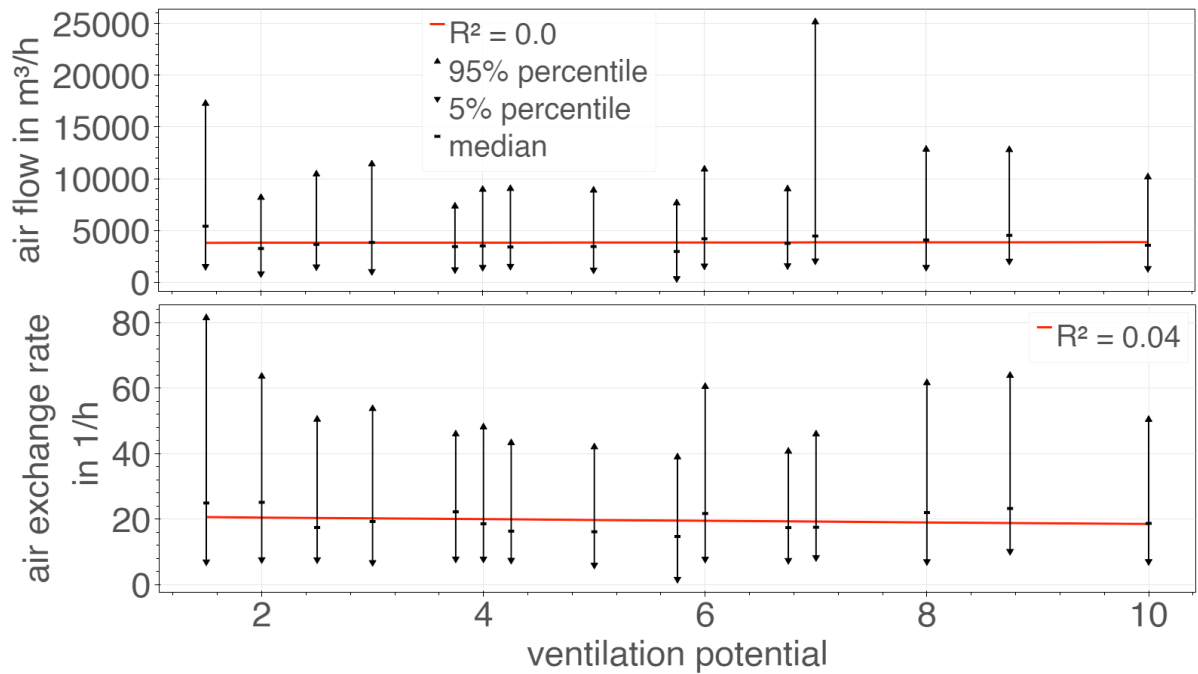

Figure 20. Air exchange rate and air flow in dependence of the ventilation potential.

It is analyzed as well, how the ventilation potential of the room influences the duration of the ventilation period. This is visualized in Figure 21. Only ventilations during times when the rooms were occupied have been evaluated and every data point is again based on at least 200 ventilations. The linear regression over median values results in a small positive correlation with $R^2$ = 0.3 and a slope of a = 0.23 min/ventilation potential. This indicates a certain trend to open more windows during ventilation, if there are more windows available. The main result is that in the majority of ventilations the ventilation potential of the room, i.e. the larger number of available windows, is not employed for faster air exchange and shorter ventilation duration.

According to Figure 4 the ventilation potential and the room volume are not proportional in the monitored rooms, but we found a large scatter of ventilation potential in average sized rooms around 200 m³ volume. But smaller rooms have smaller ventilation potential and the largest rooms have larger ventilation potential. This trend turns out to cause longer ventilation duration in the larger rooms, as shown in Figure 21.

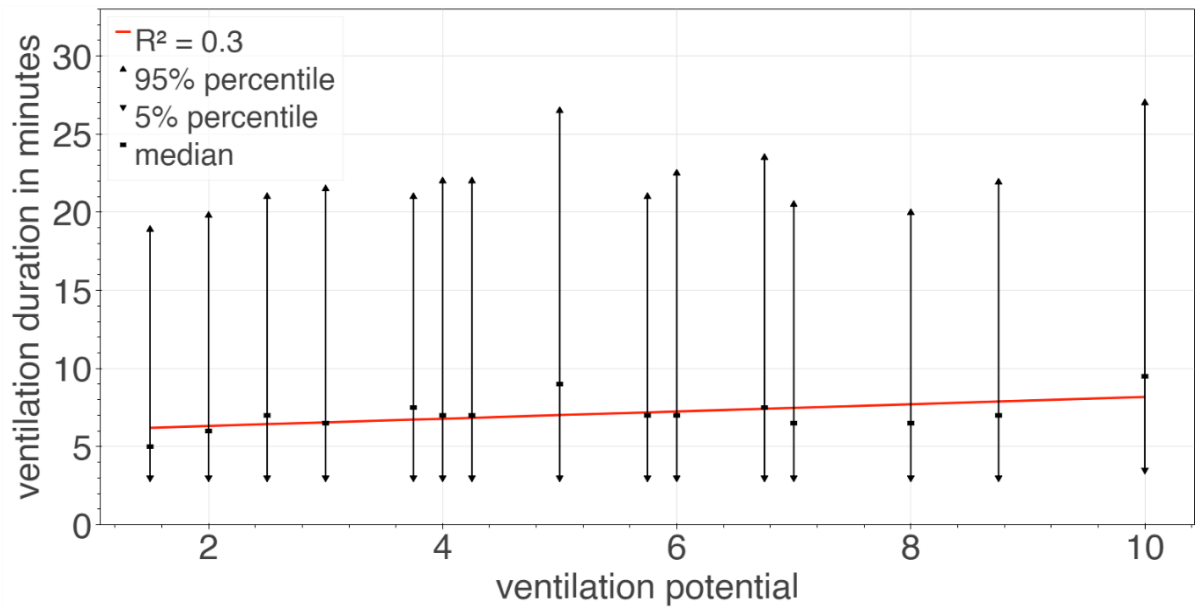

Figure 21 Ventilation duration in dependence of the ventilation potential

**Influence of room volume on air ventilation**

The influences of room volume on the air exchange rate and air flow are shown in Figure 22. Again, only ventilations in occupied rooms with at least 200 ventilations per data point were considered. Rooms with technical ventilation systems installed were excluded.

The linear regression of the median values of the air exchange rate resulted in a small negative correlation with $R^2$ = 0.29 and slope of a = -0.082 (1/h)/m³. This could mean that rooms with a higher volume get increasingly harder to ventilate as the air flow does not triple with triple room volume as would be required to get the same air exchange rate in all rooms.

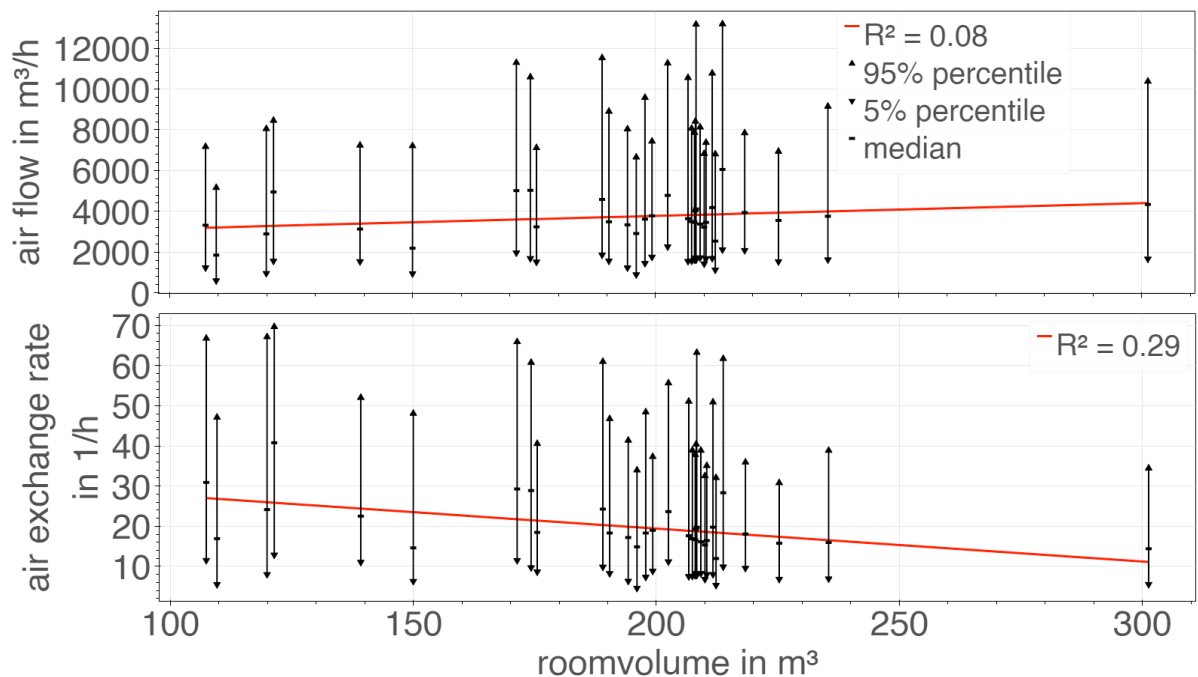

Figure 22 Influence of room volume on air exchange rate and air flow

The room volume thus also influences the ventilation duration, which is analyzed in Figure 23. Ventilations during times in occupied rooms with at least 200 ventilations per data point were evaluated. The linear regression results in very small positive correlation of $R^2 = 0.14$ with a slope of $a = 0.01$ min/m³. This result is coherent with the result of Figure 19 since smaller air exchange rates result in longer ventilations to satisfy the sensor instrument recommendation.

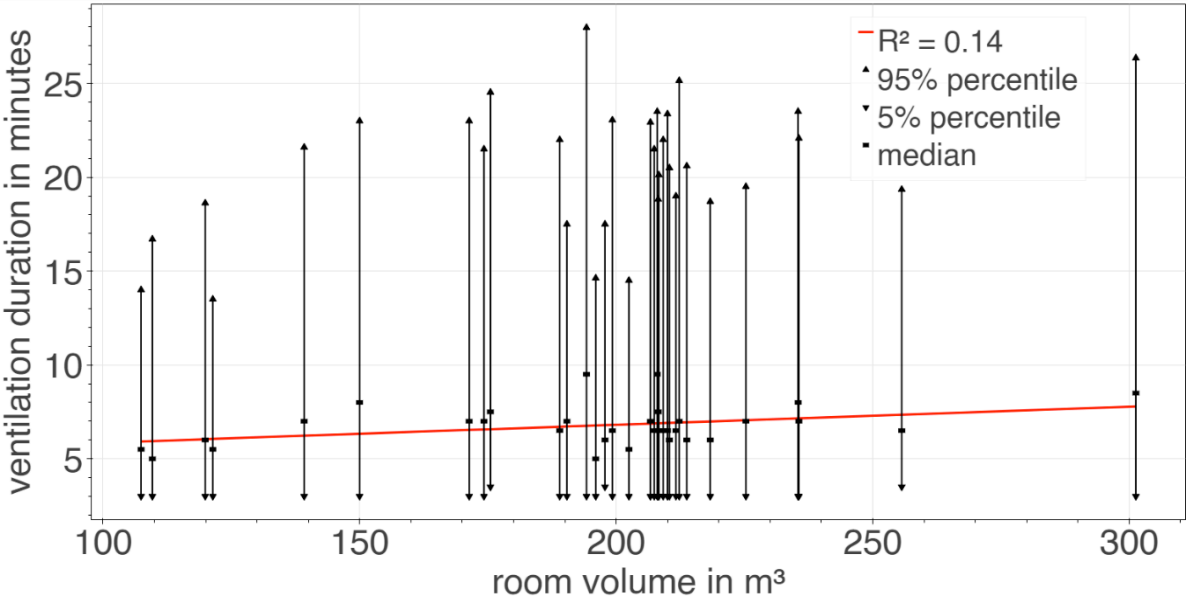

Figure 23 Ventilation duration over room volume

## Compliance overview

To evaluate the compliance to the sensor instrument "airooom" recommendations we first have to define this term. The compliance we define here as the willingness to follow the ventilation recommendation given by the instruments light signals. To measure the compliance, we use the "compliance time", which denotes the time span between the start of red phase (or yellow phase) and the start of the following ventilation. The general recommendation given to the schools is to start a ventilation process when the instrument lights turns red. After 60 sec of red light, a short "beeping" sound is additionally produced, after 2 min two short beeps are emitted and after 3 min a long beep occurs, which increases and supports the light signal effect.
We analyze the compliance time to a yellow phase as well, since we found that a lot of ventilations were performed as reaction to a yellow signal light. It has to be mentioned that there can be a time delay between the opening of the windows and the detection of a drop in $CO_2$ concentration due to air circulation within in the room. In tests this delay was measured to be around 2 minutes, but can be lower depending on the position of the sensor instrument "airooom" in relation to the opened window and the doors. Therefore, the true compliance time might be slightly lower than the number presented in the following.

Table 2 contains the 75 % and 50 % percentiles of the compliance times to the red and yellow sensor instrument light signals. We find that the compliance to red is relatively small in most cases. The yellow compliance time approximately doubles this time span. With the fact in mind that a ventilation reaction to a yellow "airooom" signal was not advised as necessary, this is a relatively short reaction time.

| percentile | compliance time red | compliance time yellow |
| --- | --- | --- |
| 75% | 4.5 minutes | 8.5 minutes |
| 50% | 2 minutes | 4 minutes |

Table 2 Compliance time percentiles

**Influence of usage time on compliance**

To test if the compliance is affected by habituation, as the duration of how long the sensor instrument "airooom" has already been in use, we evaluated the compliance time over the time in days since the "airooom" instrument had been installed in each monitored room. The daily median is shown in Figure 24 with at least 40 red/yellow phases per data point. We find no significant change in compliance over long usage times. This can be interpreted as habituation having no negative effect like ignoring the light changes during longer times of usage.

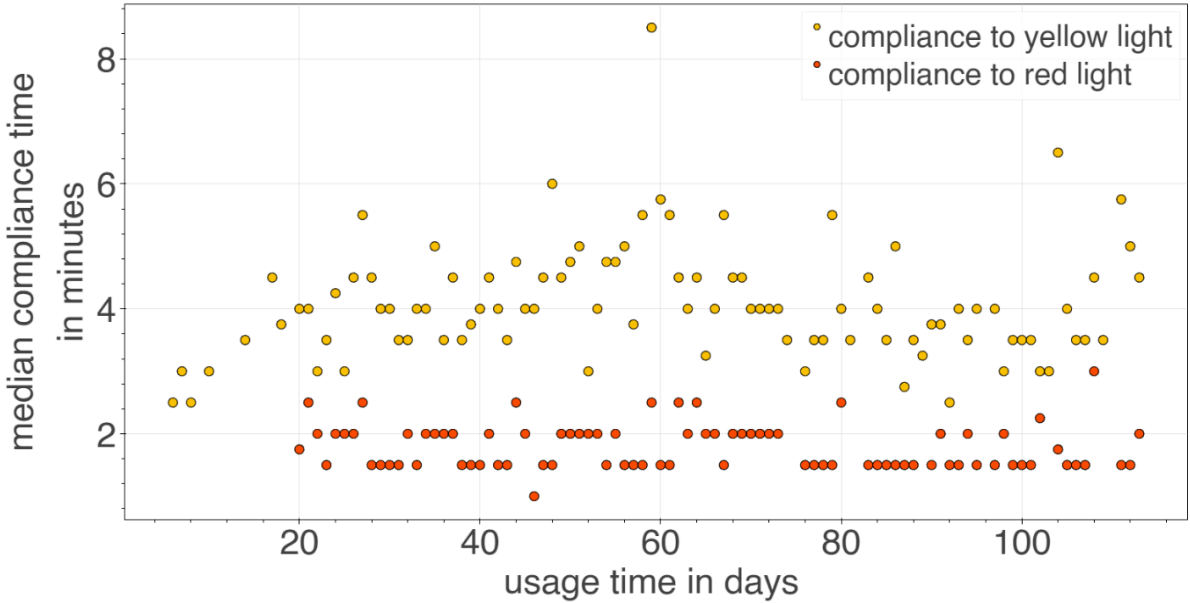

Figure 24 Daily median compliance time over the timespan since the sensor instrument "airooom" was installed.

**Seasonal influences on compliance**

The hypothesis that with lower outdoor temperatures the ventilations might get decreasingly comfortable has been investigated. This could result in lower compliance times, which we investigated and depict in Figure 25 and Figure 26. The Figures show the compliance times against the outdoor temperature in 2 °C wide bins, which contain at least 150 yellow phases in Figure 25 and 100 red phases in Figure 26.
No significant trend was identified for compliance times to "yellow" and "red" signal light, which indicates a good compliance to the sensor instrument recommendations even at lower temperatures.
The increase of compliance time for red light at higher temperatures in Figure 26 is not fully understood. Since the number of voluntary green-green ventilations increases significantly for higher temperatures the longer compliance time in Figure 26 refers to a lower number of ventilation recommendations of the sensor instrument.

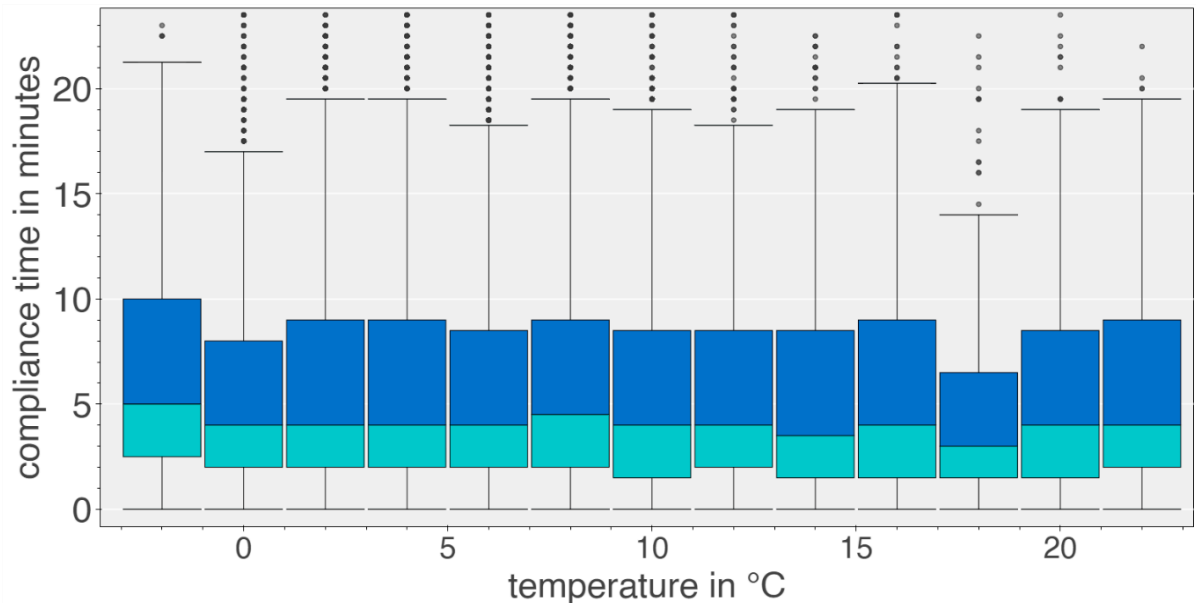

Figure 25 Boxplot of the compliance reaction time to yellow light signal depending on outdoor temperature.

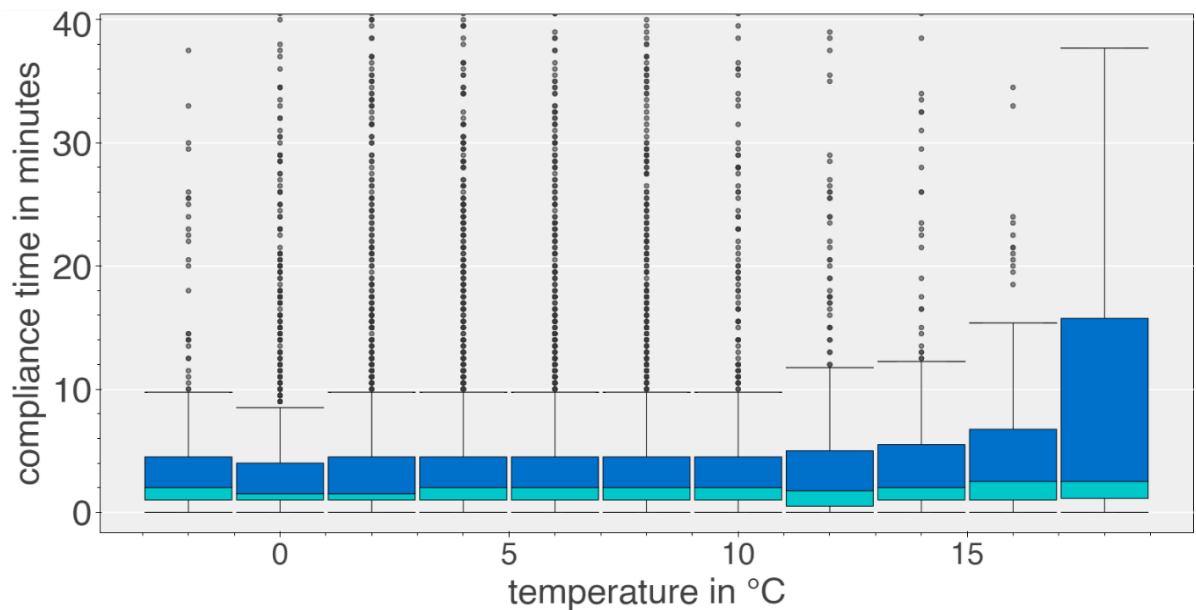

Figure 26 Boxplot of compliance reaction time to red light signal depending on outdoor temperature.

**Influence of local incidence or regional differences on compliance**

The hypothesis that with higher local incidence the compliance increases due to increased need for hygienically acceptable air has been investigated as well. We find no influence of the local incidence on the compliance.
We analyzed the recorded data as well for regional differences in compliance, but no significant effects were detected.

## 4. Conclusions

In this large study about the air ventilation in 329 classrooms in schools in Lower Saxony in the north of Germany we investigated the ventilation situation and installed in all rooms a novel instrument "airooom" consisting of a microcontroller-based data acquisition platform for $CO_2$, temperature, humidity and sound level sensors to give a ventilation recommendation based on the $CO_2$ level as a measure being correlated to aerosol emission by breathing. The data were recorded 24 hours each day and collected for 6 months since June 2021 for further evaluation. A main result of this study is that $CO_2$ monitoring for air quality control is one of very few measures against the COVID-19-pandemic, which allows quantitative, long-term, country-wide overview of effectiveness and compliance. With our visual signaling in the monitored room the air quality can reliably be kept in the hygienically acceptable range below 1000 ppm independent of outdoor temperature and for long usage of more than 6 months during summer and winter.

The compliance to a visual signaling device for $CO_2$ monitoring is very high and the achieved air ventilation by window opening in the monitored rooms is sufficient in almost all situations independent from season and outdoor temperature. In addition, critical rooms can be identified by evaluation of the stored data for further upgrading by forced ventilation or filtering.

## Acknowledgements
We gratefully acknowledge all the school members, teachers and pupils, and all members of local authorities, who contributed to the success of this study.

## Conflict of interest
The authors declare no conflict of interests. The installation and operation of the sensor instruments "airooom" have been financially supported by the Ministry of Education (MK) in Lower Saxony, Germany.

## Author contributions
Meinhard Schilling and Dean Ciric have set up the organization and technical equipment of the study and wrote most of the text. Julius Mumme has improved the "airooom" instrument. Simon Pelster, Timo Jelden and Eric Schlegel were involved in data sampling and evaluation.

## Peer Review
The peer review history of this article is available at …

## DATA AVAILABILITY STATEMENT
The data that support the findings of this study are available from the corresponding author upon reasonable request.

ORCID
Meinhard Schilling 0000-0002-7384-2028